\documentclass[letterpaper, 10 pt, conference]{ieeeconf}  
\IEEEoverridecommandlockouts                              
\usepackage{epsfig} 
\usepackage{graphicx}
\usepackage{amsmath}
\usepackage{array}
\usepackage{amssymb}
\usepackage{verbatim}
\usepackage{cleveref}
\usepackage{color}
\usepackage{relsize}
\usepackage{fancyhdr}
\usepackage{graphics}
\usepackage{setspace}
\usepackage{epstopdf}
\usepackage{multirow}
\usepackage{lipsum}
\usepackage{textcomp}
\usepackage{mathtools}
\usepackage{cite}
\usepackage{setspace}
\usepackage{booktabs}

\title{\LARGE \bf
Model Predictive Climate Control of Connected and Automated Vehicles for Improved Energy Efficiency}

\author{Hao Wang$^{1}$, Ilya Kolmanovsky$^{2}$,  Mohammad Reza Amini$^{1}$, and Jing Sun$^{1}$
\thanks{*This project is supported by the United States Department of Energy (DOE), ARPA-E NEXTCAR program (Award No.: DE-AR0000797).}
\thanks{$^{1}$Hao Wang, Mohammad Reza Amini, and Jing Sun are with the Department of Naval Architecture \& Marine Engineering, University of Michigan, Ann Arbor, MI 48109 USA. Emails: {\tt\small \{autowang, jingsun, mamini\}@umich.edu}}%
\thanks{$^{2}$Ilya Kolmanovsky is with  the Department of Aerospace Engineering, University of Michigan, Ann Arbor, MI 48109 USA. Email: {\tt\small ilya@umich.edu}}%
}

\renewcommand{\baselinestretch}{1}
\begin{document}

\maketitle

\begin{abstract}

This paper considers an application of model predictive control to automotive air conditioning (A/C) system in future connected and automated vehicles (CAVs) with battery electric or hybrid electric powertrains. A control-oriented prediction model for A/C system is proposed, identified, and validated against a higher fidelity simulation model (CoolSim). Based on the developed prediction model, a nonlinear model predictive control (NMPC) problem is formulated and solved online to minimize the energy consumption of the A/C system. Simulation results illustrate the desirable characteristics of the proposed NMPC solution such as being able to enforce physical constraints of the A/C system and maintain cabin temperature within a specified range. Moreover, it is shown that by utilizing the vehicle speed preview and through coordinated adjustment of the cabin temperature constraints, energy efficiency improvements of up to 9\% can be achieved.

\end{abstract}

\section{INTRODUCTION}

The information available through V2V, V2I, and advanced sensors in connected and automated vehicles (CAVs) provides increased situational awareness, preview of traffic conditions, and can facilitate intelligent decision-making in powertrain and vehicle control applications. Of particular interest is the improvement in fuel economy and/or the reduction in energy consumption which can be achieved in CAVs.

Thermal loads, such as those used for heating, ventilation, \& air conditioning (HVAC) of the passenger compartment, and for the electric motor and battery
package cooling, represent the most significant auxiliary loads for light-duty vehicles~\cite{Rugh2008}. It has been estimated that, in the United States, about 7 billion gallons of fuel is consumed per year just to power the air conditioning (A/C) system for light-duty vehicles~\cite{Rugh2008}. A study performed at Argonne National Lab showed a 53.7\% reduction in vehicle driving range due to air conditioning and 59.3\% reduction in vehicle driving range due to heating for Ford Focus EV, tested over the Urban Dynamometer Driving Schedule (UDDS)~\cite{Jeffers2015}. Similarly, a significant reduction of driving range was also reported for Nissan Leaf~\cite{Jeffers2015} and in a recent work by National Renewable Energy Lab \cite{NRELwhitepaper}. 

Many of the CAV related research activities, such as eco-driving and platooning, have focused on reducing traction power related losses, whereas the impact of thermal management has not been fully explored. Previous research has addressed the energy management of the A/C system for vehicles with traditional internal combustion engines (ICEs)~\cite{Rostiti15}\cite{Zhang15}, where the A/C compressor is belt-driven by the ICE. The energy management problem considered in these references was solved by a dynamic programming (DP) algorithm.  However, as vehicle power sources are becoming more electrically dominant, their energy management requires different strategies than those used for vehicles with traditional ICEs. Recognizing the pressing need to integrate advanced thermal management into overall vehicle energy optimization, a predictive climate control strategy is developed in this paper to reduce A/C system energy consumption.


The ability to enforce constraints and account for future operating conditions in rendering control decisions are among the key appealing features of MPC.
For HVAC control in buildings, predictive temperature management strategies have been studied extensively and showed substantial energy saving potentials (see \cite{Kelman11}, \cite{Oldewurtel12}, and \cite{MA12}). Similar problems for vehicles have been addressed much less and they have several distinct aspects compared to temperature control in buildings, including faster temperature dynamics, complicated passenger comfort requirements, vehicle speed dependence, and the need for solution with low computational footprint suitable for onboard implementation. 

In this paper, we are focusing on hybrid and electric vehicle applications, where the compressor of the A/C system is driven by an electric motor and draws power directly from an onboard high-voltage battery pack. We note that comprehensive modeling of such an A/C system based on the vapor compression cycle is very involved \cite{Zhang16}. Our subsequent developments exploit a high-fidelity simulation model from \cite{Kiss13}, and a simplified control-oriented model which is used for prediction and validated against our high fidelity model. A model predictive controller for the A/C system is then developed that uses the control-oriented model for prediction and minimizes energy consumption.  We also show, through a case study, that a preview of future vehicle speed profile in CAVs can be exploited to further decrease A/C system energy consumption. 

The rest of the paper is organized as follows. Details of the high-fidelity simulation model and the development of the simplified prediction model are presented in Section~\ref{sec:1}. Our design of the model predictive climate control scheme is described in Section~\ref{sec:2}. Section~\ref{sec:3} presents simulation results and demonstrates energy efficiency improvements. The conclusions are presented in Section~\ref{sec:4}.

\section{PREDICTION MODEL DEVELOPMENT}\label{sec:1}
\subsection{CoolSim High Fidelity Model and Speed Sensitivity Analysis}
\begin{figure}[b!]
	\begin{center}
		\includegraphics[width=7cm]{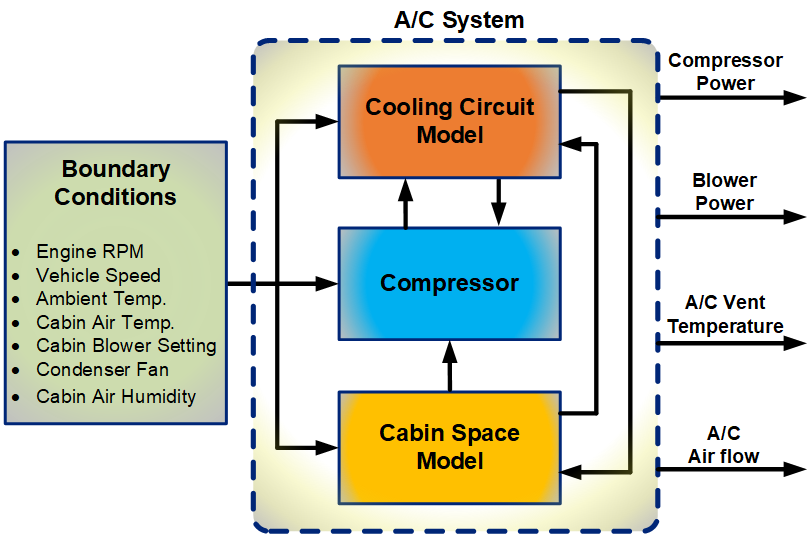}    
		\caption{Schematics of the CoolSim Simulink\textsuperscript{\textregistered} Model.}\vspace{-0.5cm}
		\label{fig:CoolSimSimulink}
	\end{center}
\end{figure}
\begin{figure}[h!]
	\begin{center}
		\includegraphics[width=6.5cm]{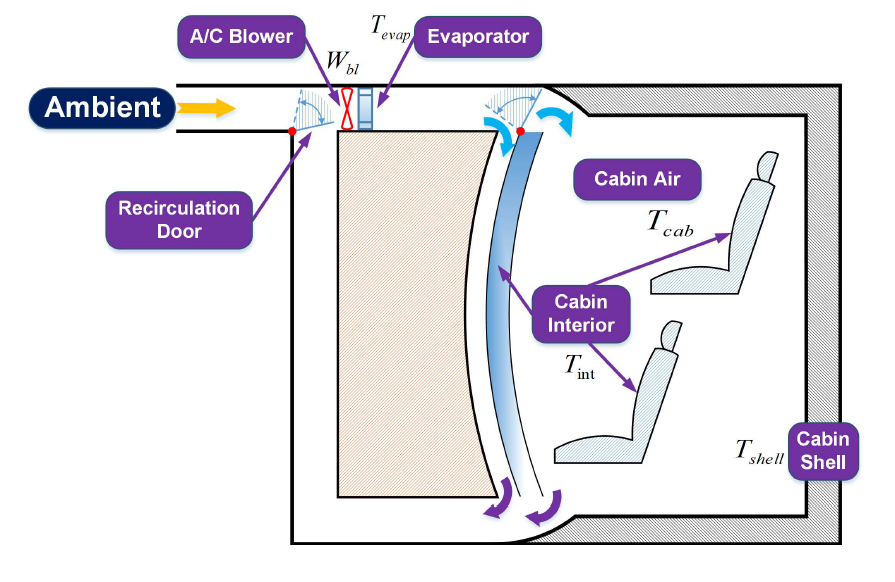}   
		\caption{Schematics of the cabin model.}\vspace{-0.75cm}
		\label{fig:CabinSchematics}
	\end{center}
\end{figure}

A high fidelity simulation model of the passenger car A/C system has been established based on CoolSim which is an open-source modeling environment available from the National Renewable Energy Lab (NREL), see \cite{CoolSim}. Fig.~\ref{fig:CoolSimSimulink} shows the schematics of the Simulink\textsuperscript{\textregistered} model of the A/C system in CoolSim. There are four major subcomponents within this model: (i) the boundary condition block, which provides the speed profiles and ambient conditions such as temperature and pressure; (ii) the cooling circuit block, which consists of detailed models of the evaporator, condenser, condenser fan, evaporator valve, and connecting pipes; (iii) the compressor block, which, as the primary energy consumer in the A/C system, is modeled separately from the cooling circuit; and (iv) the cabin space block, which models the thermal dynamics of the cabin (Fig.~\ref{fig:CabinSchematics} shows the detailed schematics and key temperatures considered in the paper). See \cite{Kiss13} for the modeling details of each subcomponent. This model is capable of simulating cycle-by-cycle behavior of the A/C system, and has been validated versus experimental data in \cite{Kiss13}. While both electric-driven and belt-driven compressor configurations are available, the electric-driven one is considered in this paper.

\begin{figure}[h!]
	\begin{center}
		\includegraphics[width=8cm]{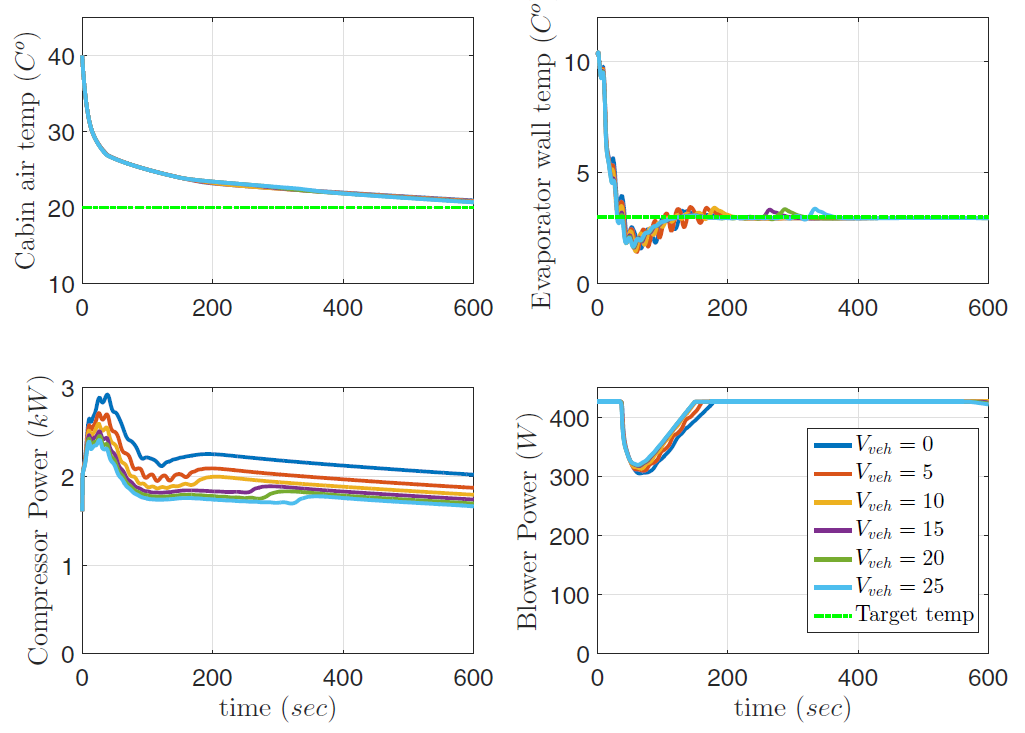} \vspace{-0.4cm}   
        \caption{Vehicle speed sensitivity demonstrated on CoolSim model.} \vspace{-0.6cm}
		\label{fig:SpeedSensitivity} 
	\end{center}
\end{figure}

The nominal controller implemented in this model consists of two Proportional-plus-Integral (PI) control loops with anti-windup and A/C on-off logic. One of the PI loops adjusts the compressor speed for tracking the evaporator wall temperature set-point. The other PI loop regulates the blower speed in order to track the cabin air temperature set-point. The recirculation rate of the cabin air depends proportionally on the difference between the cabin air temperature and the ambient temperature and is saturated according to physical feasible limits. Fig.~\ref{fig:SpeedSensitivity} gives an example of the system responses at different vehicle speeds with the nominal controller. The simulation is performed for a time period of 600 $sec$ at different constant vehicle speeds ($V_{veh}=$ 0, 5, 10, 15, 20, and 25~$m/s$) and for the same target cabin air temperature set-point. The simulation results of the CoolSim model indicate that the efficiency of the A/C system increases as the vehicle speed increases. This observation is consistent with the underlying physics, as the effective ram air speed through the condenser increases as vehicle speed increases, so that the condenser dissipates the heat faster, which leads to higher overall efficiency for the A/C system. Similar conclusion is reached in \cite{Zhang2017}.  Table~\ref{tb:ACconsump} summarizes the total energy consumption over the simulation run for different cases shown in Fig.~\ref{fig:SpeedSensitivity}. According to the values listed in the second row of Table~\ref{tb:ACconsump}, the efficiency of the A/C system increases by approximately $30\%$ as the vehicle speed increases from 0~$m/s$ (stop condition) to 25~$m/s$. The sensitivity to vehicle speed is even more pronounced if considering energy consumption normalized by the traveling distance (see the last row of Table~\ref{tb:ACconsump}). A vehicle traveling at higher speed spends less time to cover the same distance, reducing the A/C operating time and thus the associated energy consumption. This speed sensitivity can be exploited in the A/C predictive controller design. To put the numbers in Table~\ref{tb:ACconsump} in perspective, we note that the A/C energy consumption is about a third of traction power in city driving. \vspace{-0.35cm}

\begin{table}[h!]
	\centering
	\caption{Energy consumption for each case in the speed sensitivity test in Fig.~\ref{fig:SpeedSensitivity}.} 
	\label{tb:ACconsump}
    \small
    \resizebox{\columnwidth}{!}{
	\begin{tabular}{|c|c|c|c|c|c|c|} 
		\hline
		\textbf{\begin{tabular}{@{}c@{}c@{}}Vehicle\\Speed ($m/s$)\end{tabular}}   & 0 & 5& 10 & 15 & 20 & 25 \\ \hline
	    \textbf{\begin{tabular}{@{}c@{}c@{}}Energy\\Consumption\\($MJ$)\end{tabular}}  & 1.33   & 1.23    & 1.17 & 1.13 & 1.10 & 1.07 \\ \hline 
	    \textbf{\begin{tabular}{@{}c@{}c@{}}Energy\\Consumption\\($MJ/km$)\end{tabular}}  & NA   & 0.410    & 0.195 & 0.126 & 0.092 & 0.071 \\ \hline 
	\end{tabular}} \vspace{-0.5cm}
\end{table}

\subsection{A/C System Prediction Model and Its Validation}
\subsubsection{Prediction Model}
A control-oriented model for the dynamics of A/C system, which will be used as a prediction model in the implementation of MPC, is described in this section. This prediction model is motivated by physics \cite{Fayazbakhsh13} and is based on a similar approach as for building HVAC systems (see \cite{Kelman11}, and \cite{Oldewurtel12}). The model is discrete-time, has two staes ($T_{cab}$ and $T_{evap}$), and has the form,
\begin{eqnarray}
\label{eqn:1}
T_{cab}(k+1)&=&T_{cab}(k)+\gamma_1(T_{int}(k)-T_{cab}(k))\nonumber\\&+&\gamma_2(T_{shell}(k)-T_{cab}(k)) \nonumber\\&+&\gamma_3(T_{ain}(k)-T_{cab}(k))W_{bl}(k)+\tau_{1},\hspace{+0.3cm}\\
\label{eqn:2}
T_{evap}(k+1)&=&\gamma_4T_{evap}(k)\nonumber\\&+&\gamma_5(T_{evap}(k)-T_{evap,set}(k))+\tau_{2},\\
\label{eqn:3}
T_{ain}(k)&=&\gamma_6T_{evap}(k)+\gamma_7W_{bl}(k)+\tau_{3}.
\end{eqnarray} 
In (\ref{eqn:1})-(\ref{eqn:3}), $T_{cab}$, $T_{int}$, $T_{shell}$, $T_{evap}$, and $T_{ain}$ represent the temperatures (in $C^o$) of the cabin air, the cabin interior (e.g. seats and panels), the cabin shell, the evaporator wall and the cabin inlet air flow, respectively. The control inputs to the model are $W_{bl}$ (blower flow rate in $kg/s$) and $T_{evap,set}$ (evaporator wall temperature set-point in $C^o$). The model parameters, $\gamma_i~(i=1,2,...,7)$ and $\tau_j~(j=1,2,3)$ are identified from CoolSim model. Note that the model given by (\ref{eqn:1})-(\ref{eqn:3}) is nonlinear due to a bilinear term in (\ref{eqn:1}). The structure of the model reflects the following assumptions:
\begin{enumerate}
	\item The recirculation rate of the cabin air ($\alpha_{recirc}$) is constant ($\alpha_{recirc}\in\left[0~1\right]$, where $\alpha_{recirc}=0$ means cabin inlet air is all from ambient while $\alpha_{recirc}=1$ means all cabin air is recirculated via A/C system).
	\item The dynamics of $T_{int}$ and $T_{shell}$ are slower than the dynamics of $T_{cab}$ and $T_{evap}$. Thus, $T_{int}$ and $T_{shell}$ are treated as measured inputs.
	\item The sensitivity of the states to vehicle speed is not reflected in the prediction model; accounting for this sensitivity is left to future research.
	\item Blower dynamics can be ignored because of its small time constant.
\end{enumerate}

\subsubsection{Model Identification}
Next, the outputs from the CoolSim model excited with random input signals are sampled at $0.2$Hz to generate data for identifying the unknown parameters in (\ref{eqn:1})-(\ref{eqn:3}). The resulting identified parameters are $\gamma=\left[\gamma_1~\gamma_2~ ... ~\gamma_7\right]=\left[0.2451~0.0867~1.2999~1.0047~-0.5176~0.4553~34.9579\right]$ and $\tau=\left[\tau_1~\tau_2~\tau_3\right]=\left[-0.1842~-1.3226~154.4995\right]$.

\subsubsection{Model Validation}
\begin{figure}[h!]
	\begin{center}
		\includegraphics[width=8cm]{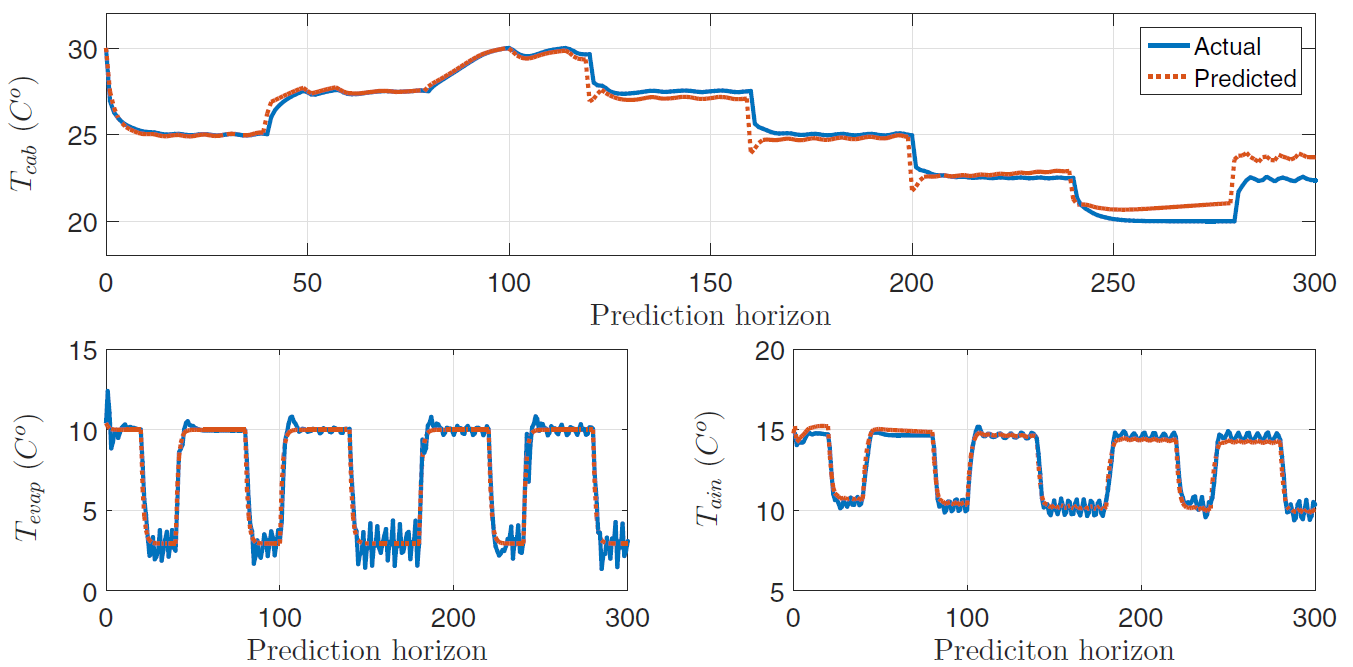} \vspace{-0.3cm}     
		\caption{Model predictions initialized at a specific time point.}\vspace{-0.5cm}     
		\label{fig:Prediction1}
	\end{center} \vspace{-0.35cm}
\end{figure}

\begin{figure}[h!]
	\begin{center}
		\includegraphics[width=8cm]{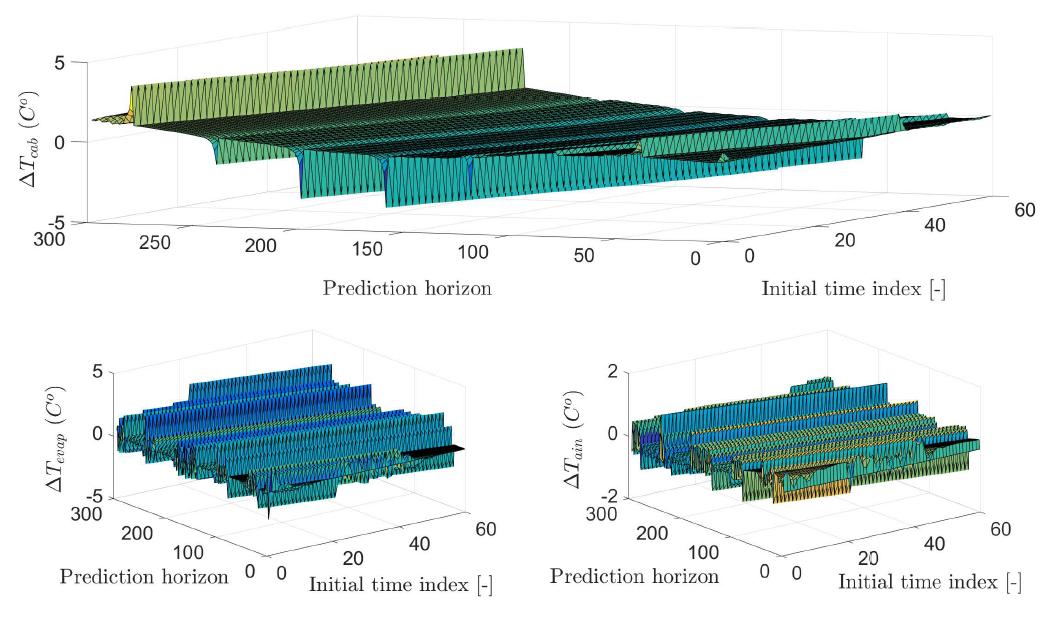}\vspace{-0.3cm}     
		\caption{Model prediction errors for different initial time instants.} \vspace{-0.5cm} 
		\label{fig:PredictionError1} 
	\end{center} 
\end{figure}

Fig.~\ref{fig:Prediction1} shows the validation results of the control-oriented model, which predicts the system behaviors over $300$ steps into the future ($1500$ $sec$) given the measurements only at the initial time step. $T_{int}$ and $T_{shell}$, which are external inputs to the model, are assumed to be constant over the prediction horizon. By comparing the behavior of the control-oriented model with the high fidelity CoolSim model in Fig.~\ref{fig:Prediction1}, it can be seen that the identified model ((\ref{eqn:1})-(\ref{eqn:3})) provides reasonably accurate results. In order to further confirm the accuracy of the control-oriented model, predictions are performed at $60$ different initial time instants. As shown in Fig.~\ref{fig:PredictionError1}, the prediction error, which is the difference between the predicted and actual values, is bounded for the system's states  ($T_{cab}$ and $T_{evap}$) and output ($T_{ain}$) within a couple of degrees for most of the time. \vspace{-0.25cm}

\section{Model Predictive Climate Control Formulation}\label{sec:2} 
In this section, a nonlinear MPC (NMPC) optimization problem is formulated for the A/C system in which, the objective is to minimize the energy consumption of the A/C system during vehicle operation. Two major energy consumers in the A/C system are the compressor and the blower. According to \cite{Kelman11}, their consumed powers can be estimated by:
\begin{eqnarray}
\label{eqn:Pc}
P_c&=&\frac{c_p}{\eta_{cop}}W_{bl}(T_{amb}-T_{ain})\nonumber\\
&=&\frac{c_p}{\eta_{cop}}W_{bl}(T_{amb}-\gamma_6T_{evap}-\gamma_7W_{bl}-\tau_3),\\
\label{eqn:Pbl}
P_{bl}&=&\beta_1 W_{bl}^2+\beta_2 W_{bl}+\beta_3,
\end{eqnarray}
where $P_c$ and $P_{bl}$ represent the powers of the compressor and blower, respectively, $c_p$ is the specific heat capacity of air at constant pressure, $\eta_{cop}$ is the coefficient of performance (COP) of the A/C system \cite{Bhatti97}, and $\left[\beta_1~\beta_2~\beta_3\right]=\left[24156~-1974.2~49.318\right]$ are the parameters identified from CoolSim data.

Next, we define
\begin{eqnarray} 
\label{eqn:definevar}
x=\left[\begin{array}{c}
T_{cab} \\ 
T_{evap}
\end{array} \right],u=\left[\begin{array}{c}
W_{bl} \\ 
T_{evap,set}
\end{array} \right],v=\left[\begin{array}{c}
T_{int} \\ 
T_{shell}\\
T_{ain}
\end{array} \right]
\end{eqnarray}
and we let $x(i|k), u(i|k), v(i|k)$ denote the predicted values of the states, inputs and auxiliary variables, respectively, at time $k+iT_s$, with the prediction made at the time instant $k$.

Based on the proposed model (\ref{eqn:1})-(\ref{eqn:3}), the electrical power consumption model (\ref{eqn:Pc})-(\ref{eqn:Pbl}), and definitions (\ref{eqn:definevar}), the NMPC optimization problem is defined as 
\begin{eqnarray}
\label{costfun}
\min\limits_{U(k)}~~\sum_{i=0}^{N_p} P_c(i|k)+P_{bl}(i|k)+\sum_{i=0}^{N_c} a_{sl} {v_{sl}(i|k)},
\end{eqnarray}
\begin{eqnarray}
\label{eqn:nonlinearconstr}
&s.t. ~~ \left[\begin{array}{c}
x(i+1|k) \\ 
x(i|k) \\ 
u(i|k)\\
v(i|k)
\end{array} \right]^\top C~~\left[\begin{array}{c}
x(i+1|k) \\ 
x(i|k) \\ 
u(i|k)\\
v(i|k)
\end{array} \right]\nonumber \\ &+ A_1\left[\begin{array}{c}
x(i+1|k) \\ 
x(i|k) \\ 
u(i|k)\\
v(i|k)
\end{array} \right]+\tau_1=0,i=0,...,N_p,\\
\label{linearconstr}
&A_2\left[\begin{array}{c}
x(i+1|k) \\ 
x(i|k) \\ 
u(i|k)\\
v(i|k)
\end{array} \right]+\left[\begin{array}{c}
\tau_2 \\ 
\tau_3 \end{array} \right]=0,i=0,...,N_p,
\end{eqnarray}
\begin{eqnarray}
\label{stateconstr}
x(i|k)\geq\underline{x}(i|k)-v_{sl}(i|k), i=0,...,N_c,\\
x(i|k)\leq \overline{x}(i|k)+v_{sl}(i|k), i=0,...,N_c,\\
v_{sl}(i|k)\geq 0, i=0,...,N_c, \\
\label{inputconstr}
\underline{u}(i|k)\leq  u(i|k)\leq \overline{u}(i|k), i=0,...,N_u-1,\\
x(0|k)=x(k),\\ u(0|k)=u(k), \label{eqn:ini_constr}
\end{eqnarray}
where $U(k)=(u^{\top}(0|k),...,u^{\top}(N_u-1|k))$ and $u^{\top}(i|k)=\left[T_{evap,set}(i|k),  W_{bl}(i|k)\right]^\top$, $N_p$ is the prediction horizon, $N_u\leq N_p$ is the control horizon, and $N_c\leq N_p$ is the constraint horizon. As in other practical applications of MPC, the state constraints are relaxed with slack variables to avoid infeasibility due to model mismatch. The vector $v_{sl}\in \mathbb{R}^{2\times 1}$ represents the slack variable vector and $a_{sl}=\left[10^5~~10^5\right]$ is the penalty on the slack variables. Equality constraints (\ref{eqn:nonlinearconstr}) and (\ref{linearconstr}) are informed by the system dynamics (\ref{eqn:1})-(\ref{eqn:3}), where $C\in \mathbb{R}^{9\times 9}$, $A_1\in \mathbb{R}^{1\times 9}$, and $A_2\in \mathbb{R}^{2\times 9}$ are constant matrices. In particular, $C(i|k)$ is symmetric and indefinite. The lower and upper bounds on the states are given, respectively, by $\underline{x}(i|k)$ and $\overline{x}(i|k)$. The lower and upper bounds on the inputs are given, respectively, by $\underline{u}(i|k)$ and $\overline{u}(i|k)$.

Note that additional state constraints, such as the constraint on the time rate of change of the cabin temperature and others \cite{Diana14}, can be similarly introduced to ensure passenger comfort. These additional constraints will be considered in future work. The nonlinear and nonconvex optimization problem in (\ref{costfun})-(\ref{eqn:ini_constr}) is solved numerically using MPCTools package \cite{mpctools}. 

\section {MPC Simulation Results} \label{sec:3}
\subsection{Controller Implementation in Simulink\textsuperscript{\textregistered}}
\begin{figure*}
	\begin{center}
		\includegraphics[scale=0.4]{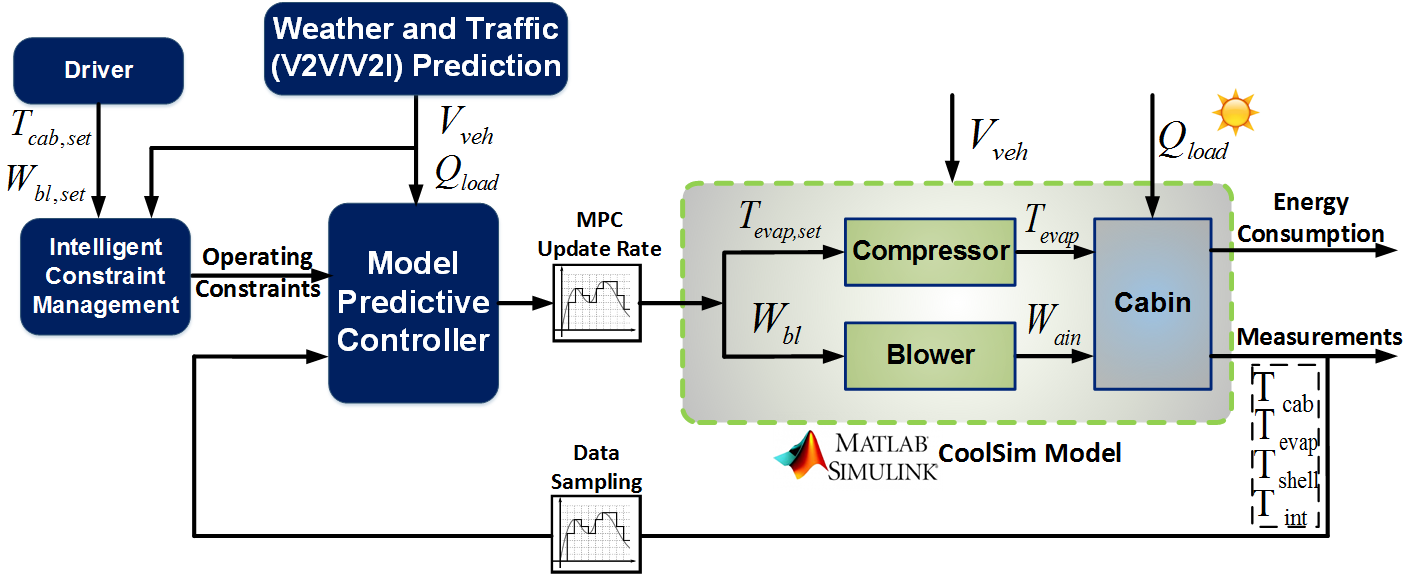}    
		\caption{Schematics of the model predictive climate control system. The Coolsim model is executed in Simulink\textsuperscript{\textregistered}, while the controller is designed in MATLAB\textsuperscript{\textregistered}, and called by the model via an Interpreted MATLAB Function in Simulink\textsuperscript{\textregistered} environment.} 
		\label{fig:ACcontroldiagram}
	\end{center}
\end{figure*}

\begin{table}[b!]
	\caption{Constant parameters and constraints applied for implementing the NMPC} 
	\centering
	\label{tb:parameters}
	\small
	\begin{tabular}{@{}ccc@{}}
		\toprule
		\textbf{Parameter} & \textbf{Value}   & \textbf{Units} \\ \midrule
		$N_p=N_c=N_u$      & 6                & dimensionless  \\
		$\eta_{cop}$           & 3.5                & dimensionless  \\ 
		$c_p$           & 1008              & $J/(kg\cdot K)$  \\
		$T_{amb}$      	& 30                & $C^o$  \\ 
		$\overline{T}_{evap}$           & 12              & $C^o$  \\
		$\overline{W}_{bl}$           & 0.15              & $kg/s$  \\
		$\overline{T}_{evap,set}$           & 10              & $C^o$  \\
		$\underline{T}_{cab}$           & 20              & $C^o$  \\
		$\underline{T}_{evap}$           & 0              & $C^o$  \\		
		$\underline{W}_{bl}$           & 0.05              & $kg/s$  \\	
		$\underline{T}_{evap,set}$           & 3              & $C^o$  \\		\bottomrule
	\end{tabular}
\end{table}

The overall schematics of the proposed predictive climate control system are shown in Fig.~\ref{fig:ACcontroldiagram}. The NMPC controller uses sensor measurements and predictions of the weather and traffic conditions as inputs, and computes the control commands for the system to maintain the cabin temperature within the predefined comfort zone. The constraints are updated online by an Intelligent Constraint Management (ICM) block. 

In our simulations, the NMPC is implemented as an Interpreted MATLAB Function block in Simulink\textsuperscript{\textregistered}. CoolSim model is simulated with a fixed time step of 0.01~\textit{sec} to emulate the continuous-time behavior of the A/C system while a larger update interval (5~\textit{sec}) is chosen for the NMPC; thus the feedback signals from the CoolSim model are also sampled every 5~\textit{sec}. Rate transition blocks, shown in Fig.~\ref{fig:ACcontroldiagram}, are used to handle the difference in the sampling times of the CoolSim model and the NMPC controller. Table~\ref{tb:parameters} lists the constant parameters and constraints used in our simulation case studies. The constraints on the states and inputs are obtained from the operating limits of CoolSim model. The upper bound of the cabin temperature ($\overline{T}_{cab}$) is the only time-varying constraint that is updated online in our current implementation.

\subsection{Control Performance Evaluation}

\begin{figure}[h!]
	\begin{center}
		\includegraphics[width=9cm]{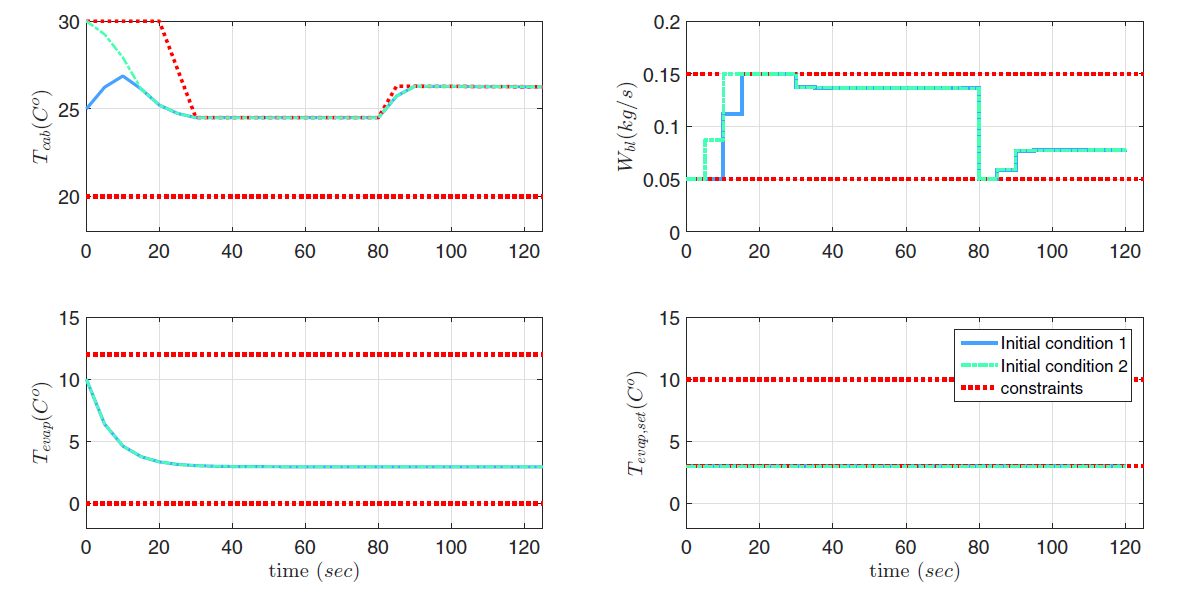}   
		\caption{Simulation results of NMPC in closed-loop with the control-oriented model.} 
		\label{fig:MPC_SimpleModel} 
	\end{center}
\end{figure}

We first implement the proposed predictive controller on the control-oriented A/C system model to assure that the design objectives are achieved. Fig.~\ref{fig:MPC_SimpleModel} shows an example of controlling the system over a $125$ $sec$ time interval with a time-varying upper limit on the cabin temperature ($\overline{T}_{cab}$) and for two choices of initial conditions to emulate a summer cabin cool-down scenario. As can be observed from Fig.~\ref{fig:MPC_SimpleModel}, all the constraints, which are represented by red dotted lines, are satisfied during the simulation period. Since we are minimizing the energy consumption of the system, the cabin temperature approaches the upper limits in steady state as expected. In this simulation, $T_{shell}$ and $T_{int}$ are set to constant values. The simulations were carried out on a laptop computer with an @2.20 GHz processor. The average time required for control computation was less than one eights of the sampling period. 

%

\begin{figure}[h!]
	\begin{center}
    		\includegraphics[width=9cm]{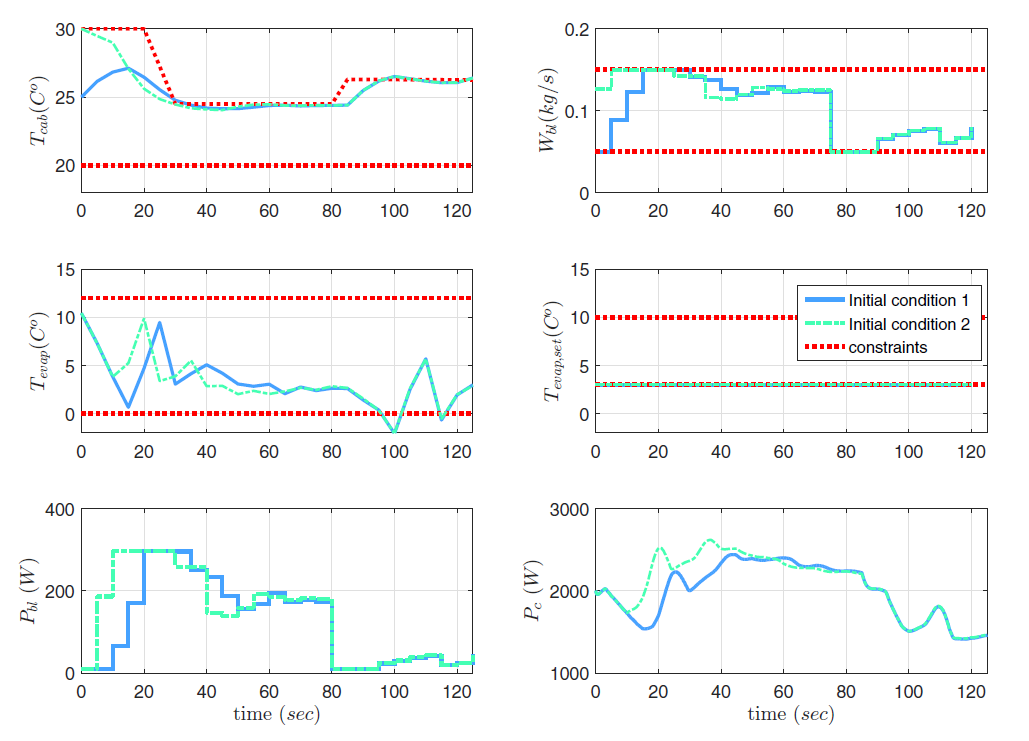} 
		\caption{Simulation results of NMPC in closed-loop with CoolSim model.} 
		\label{fig:MPC_CoolSim} 
	\end{center}
\end{figure}

Next, the same scenarios are simulated with the high fidelity CoolSim model. The system responses for two cases with different initial conditions are shown in Fig.~\ref{fig:MPC_CoolSim}. As observed from the results, the constraints are satisfied during most of the simulation period, with slight violations at some time instants. The behavior of the evaporator wall temperature deviates from our first-order approximation in (\ref{eqn:2}), especially during the transients (e.g., when there is a change in the upper limit of the desired cabin temperature). This deviation is due to model mismatch between the simplified control-oriented model and the CoolSim high-fidelity model during the transients. The first-order approximation is more accurate when the blower mass flow rate is varying slowly (e.g., in the case shown in Fig.~\ref{fig:Prediction1}). Two other CoolSim outputs, blower power and compressor power, are also reported in Fig.~\ref{fig:MPC_CoolSim}, based on which the A/C system energy consumption can be computed. In this case, $T_{shell}$ and $T_{int}$ are time-varying parameters updated by the CoolSim model. Note that the MPC controller accounts for the preview of the time-varying constraints over the prediction horizon. 
\vspace{-0.4cm}

\subsection{NMPC Implementation for CAVs}
\begin{figure}[h!]
	\begin{center}
		\includegraphics[width=8cm]{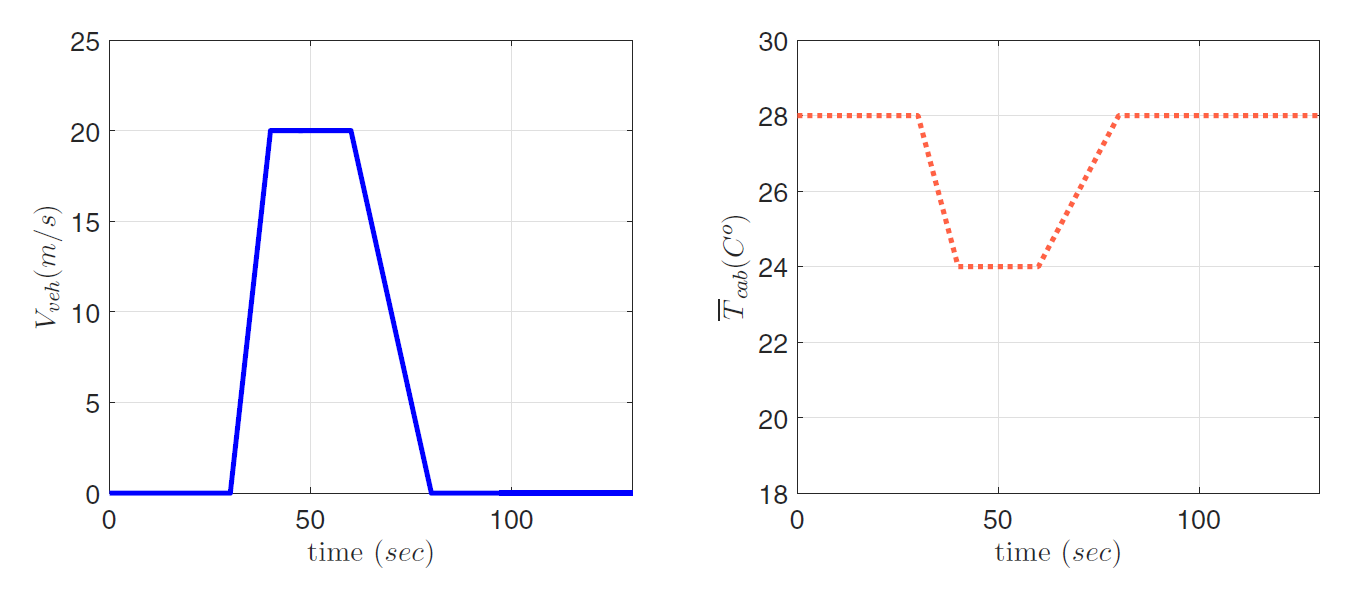} \vspace{-0.3cm}
		\caption{Assumed future speed profile and coordinated adjustment of upper bound on $T_{cab}$.} 
		\label{fig:MPC_speed_sensitivity_prep} 
	\end{center}
\end{figure}

From the speed sensitivity tests shown in Fig.~\ref{fig:SpeedSensitivity}, it is observed that the efficiency of the A/C system increases as the vehicle speed increases. In this section, an example of leveraging the predicted vehicle speed profile for reducing energy consumption by the A/C system is presented. Such a predicted vehicle speed profile is expected to be available in CAV applications; it is expected that it will be generated using V2V/V2I communications and traffic models. In our simulations, a vehicle speed profile is defined in Fig.~\ref{fig:MPC_speed_sensitivity_prep} to emulate a stop-and-go scenario; this scenario is assumed to be known at $t=0$ $sec$. According to this speed profile, and the relationship between the A/C system efficiency and the vehicle speed, a time-varying upper bound on cabin temperature, $\bar{T}_{cab}$, shown in Fig.~\ref{fig:MPC_speed_sensitivity_prep}, is defined to emphasize cooling when the vehicle speed and efficiency of A/C system are higher. 

\begin{figure}[h!]
	\begin{center}
		\includegraphics[width=8.5cm]{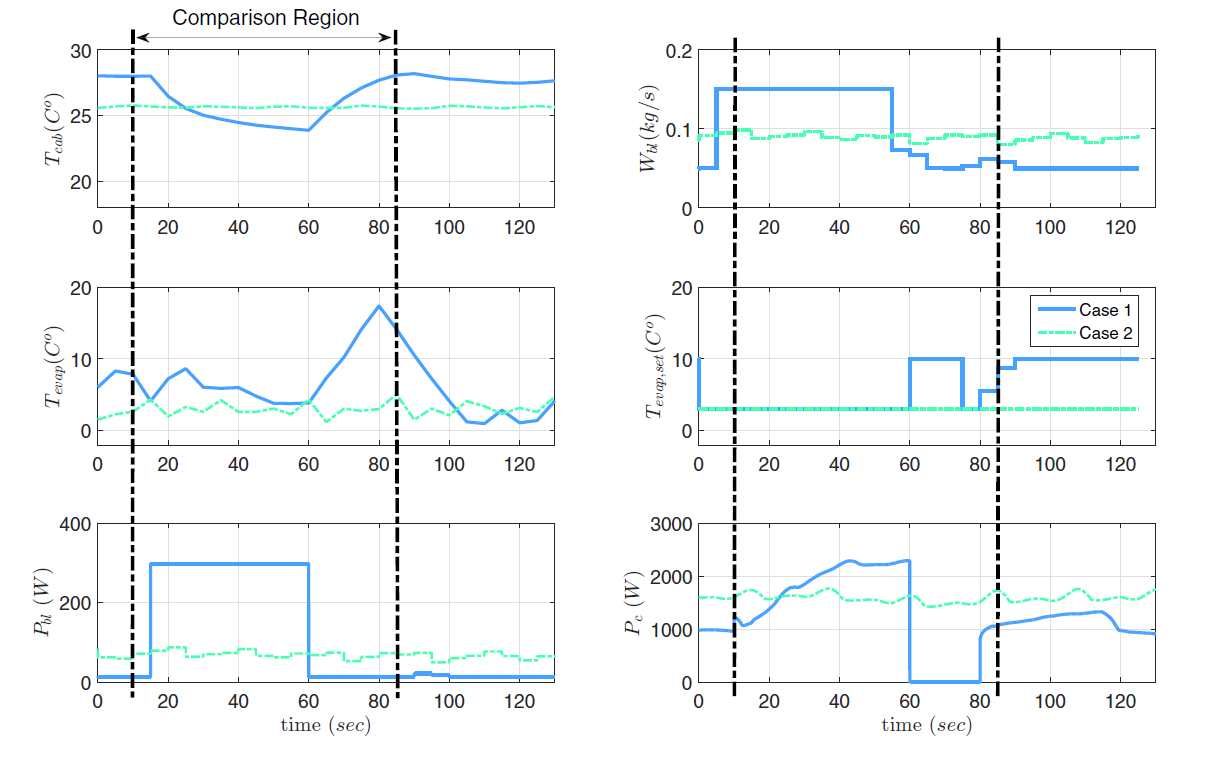} \vspace{-0.3cm}
		\caption{Comparison between two control scenarios (Case 1: coordinating the A/C system operation with predicted vehicle speed, Case 2: tracking constant cabin temperature set-point).}  
		\label{fig:MPC_speed_sensitivity} 
	\end{center}
\end{figure}

Two cases are shown in Fig.~\ref{fig:MPC_speed_sensitivity} to demonstrate the energy saving solutions.  The simulation results of testing the predictive climate controller based on the designed cabin temperature upper bound are shown in Fig.~\ref{fig:MPC_speed_sensitivity} as Case 1. For this case, an open-loop compressor shut-off control is implemented during the temperature recovering period (from $60~sec$ to $80~sec$). In Case 2, which represents the conventional A/C system control scheme, the controller tracks a constant cabin temperature set-point (i.e. a constant $\bar{T}_{cab}$ is used by the NMPC controller). This set-point is the average of the temperatures within the comparison region (from $10~sec$ to $85~sec$) of Case 1 shown in Fig.~\ref{fig:MPC_speed_sensitivity}. The total energy consumptions during the comparison region for Case 1 and Case 2 are $0.113$ $MJ$ and $0.124$ $MJ$, respectively. As can be seen, $9\%$ energy saving is achieved in this case study. Thus, allowing the temperature to vary within a certain passenger comfort range, and optimizing the A/C system operation for the future vehicle speed profile provides additional opportunities to improve energy efficiency, compared to tracking a constant cabin temperature set-point.

\vspace{-0.3cm}

\vspace{-0.3cm}

\section{CONCLUSIONS}\label{sec:4}
A model predictive climate control framework to enable energy savings and (potentially) range extension in hybrid and electric passenger cars was presented in this paper. Exploiting opportunities emerging in connected and automated vehicles, a preview of vehicle speed and weather conditions can be integrated into HVAC control. In this work, a high fidelity A/C system model (CoolSim) was adopted as the virtual testbed for testing the predictive climate control algorithm. In order to conduct real-time optimization, a control-oriented prediction model of the A/C system was developed and validated against data from CoolSim model. Then, a nonlinear MPC (NMPC) problem was formulated and solved for minimizing the energy consumption of the A/C system. The performance of the proposed NMPC controller was validated in closed-loop with CoolSim model. In order to demonstrate the benefits of incorporating the future vehicle speed information into the A/C control problem, a speed-correlated test scenario with a time-varying cabin temperature constraint was compared with a conventional constant cabin temperature setpoint scenario. The results showed that coordinating the cabin temperature with the vehicle speed profile can result in up to 9\% improvement in the energy efficiency of the A/C system. Future work will address enhancing the prediction model to include vehicle speed sensitivity, handling additional comfort constraints, incorporating weather/thermal input forecasts, and co-optimization of vehicle speed and A/C system operation.






\end{document}